\documentclass[sigconf]{acmart}
\AtBeginDocument{%
  }

\usepackage{pifont}
\usepackage{placeins}
\usepackage{float}
\usepackage[most]{tcolorbox}
\usepackage{setspace}
\usepackage{hyphenat}
\usepackage{etoolbox}
\newcommand{\cmark}{\textcolor{green!60!black}{\ding{51}}}
\newcommand{\xmark}{\textcolor{red!80!black}{\ding{55}}}

\definecolor{metafg}{HTML}{1C2B33}
\definecolor{metabg}{HTML}{F1F4F7}
\definecolor{metablue}{HTML}{0064E0}
\DeclareFontFamily{T1}{optimistic}{}
\DeclareFontShape{T1}{optimistic}{m}{n}{<-> s * [0.88] assets/optimistic}{}
\DeclareFontShape{T1}{optimistic}{b}{n}{<-> s * [0.88] assets/optimistic}{}
\pdfmapline{+optimistic < assets/Optimistic.ttf <T1-WGL4.enc}
\newcommand{\optimisticfamily}{\fontfamily{optimistic}\selectfont}
\newcommand{\metaromanfamily}{\fontfamily{cmr}\selectfont}

\newcommand{\MetaPaperTitle}{An Event is Worth One Token: Event Tokenization for Industrial-scale LLM Recommendation}
\newcommand{\metaauthor}[1]{#1}
\newcommand{\metacoreauthor}[1]{#1$^{*}$}
\newcommand{\MetaPaperAuthors}{%
  Fan Xia$^{*,\dagger}$,
  \metacoreauthor{Zhaoheng Zheng},
  \metacoreauthor{Iman Setayesh},
  \metacoreauthor{Ruogu Lin},
  \metacoreauthor{Yiqin Pan},
  \metacoreauthor{Samarth Mittal},
  \metaauthor{Wentao Bao},
  \metaauthor{Vinti Pandey},
  \metaauthor{Sachin Patil},
  \metaauthor{Jianpeng Cheng},
  \metaauthor{Jun Xiao},
  \metaauthor{Zhuang Wang},
  \metaauthor{Xiangjun Fan},
  \metaauthor{Sri Reddy},
  \metaauthor{Minghai Chen}}

\newcommand{\metatitlebox}{%
  \tcbset{enhanced,frame hidden}
  \tcbset{left=0.5cm}
  \tcbset{right=0.5cm}
  \tcbset{top=0.5cm}
  \tcbset{bottom=0.5cm}
  \tcbset{arc=10pt}
  \tcbset{colback=metabg}
  \tcbset{before skip=0pt}
  \tcbset{grow to left by=1.5pt}
  \tcbset{grow to right by=1.5pt}
  \tcbset{overlay={\node[
    anchor=south east,
    at=(frame.south east),
    xshift=-0.5cm,
    yshift=0.5cm] {\includegraphics[width=1.5cm]{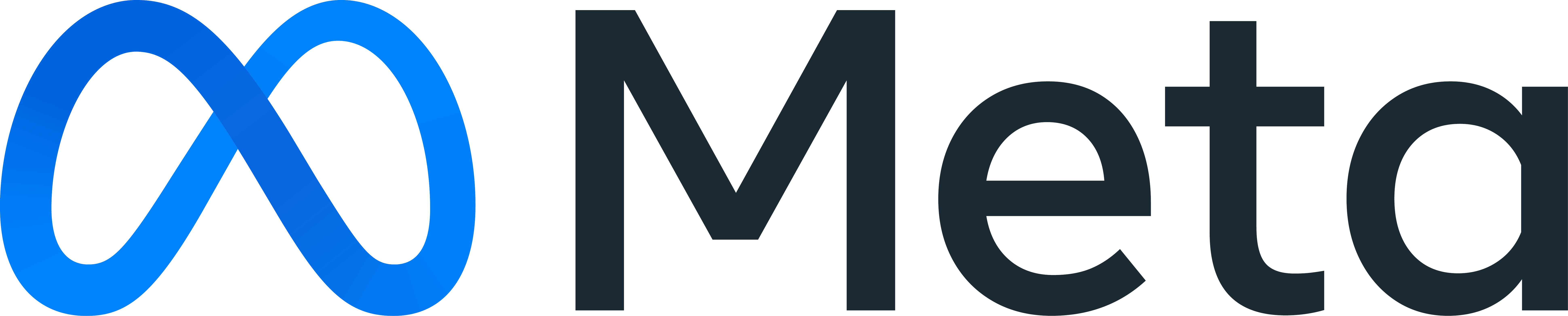}};}}
  \begin{tcolorbox}
    \setlength{\parindent}{0cm}
    \setlength{\parskip}{0.5cm}
    \fontsize{10pt}{12pt}\selectfont
    {
      \setlength{\parskip}{0cm}
      \raggedright
      \nohyphens
      {
        \setstretch{1.618}
        {\fontsize{20.74pt}{25pt}\selectfont\optimisticfamily\bfseries \MetaPaperTitle}\par
      }
      \vskip 0.2cm
      {\optimisticfamily\bfseries \MetaPaperAuthors}\par
      \vskip 0.2cm
      {\small\metaromanfamily $^{*}$Core contributors,\quad
        $^{\dagger}$Correspondence,\quad AI at Meta}\par
    }
    {\color{metafg}\fontsize{10pt}{12pt}\selectfont\metaromanfamily \MetaPaperAbstract}\par
    \vskip 0.5cm
    {\setlength{\parskip}{0cm}\small
      {\optimisticfamily\bfseries Date:} August 24, 2026\par
      {\optimisticfamily\bfseries Correspondence:} Fan Xia at
      \textcolor{metablue}{\href{mailto:xiafanaifb@meta.com}{\texttt{xiafanaifb@meta.com}}}\par}
  \end{tcolorbox}
  \tcbset{reset}
  \FloatBarrier
}

\newcommand{\makemetatitle}{%
  \twocolumn[\metatitlebox\vskip 0.38cm]
}

\setcopyright{none}
\renewcommand\footnotetextcopyrightpermission[1]{}

\begin{document}

\title{\MetaPaperTitle}

\author{Fan Xia et al.}
\affiliation{%
  \institution{AI at Meta}
  \country{United States}
}

\renewcommand{\shortauthors}{Fan Xia et al.}

\newcommand{\MetaPaperAbstract}{%
LLM-based recommendation has scaled along model capacity and sequence length, yet each position encodes only text, semantic IDs, or a few categorical features---discarding rich user, item, context, and outcome signals available at each event. Under autoregressive modeling, this yields weak queries at each position and, since each position becomes context for the next, the degradation compounds across the sequence. We propose an event-centric paradigm that represents each interaction by its full temporal snapshot, and identify a new scaling dimension we term \textbf{snapshot resolution}: the amount of information encoded per event. To efficiently scale snapshot resolution, we introduce \textbf{AMBER} (Autoregressive Modeling via Bottlenecked Event Representation), which compresses each temporal snapshot into a compact \textbf{Event Token}, a new LLM input modality. The representation is learned end-to-end, while Event Tokens are pre-computed and cached for serving, decoupling snapshot resolution from real-time serving compute. On industrial-scale ranking and retrieval benchmarks, AMBER advances the compute-quality Pareto frontier relative to alternative recommendation paradigms. At sufficient capacity, a single unified tokenizer even outperforms dedicated per-entity tokenizers, demonstrating positive transfer across structurally different entity types.
AMBER's Event Tokens also transfer across model architectures: when integrated into a heavily optimized non-LLM ranker as serving-time historical features, they yield statistically significant improvements. Further scaling Event Tokenizer capacity provides additional improvements.
}

\ccsdesc[500]{Information systems~Recommender systems}

\makemetatitle

\begin{figure}[t]
\centering
\includegraphics[width=\linewidth]{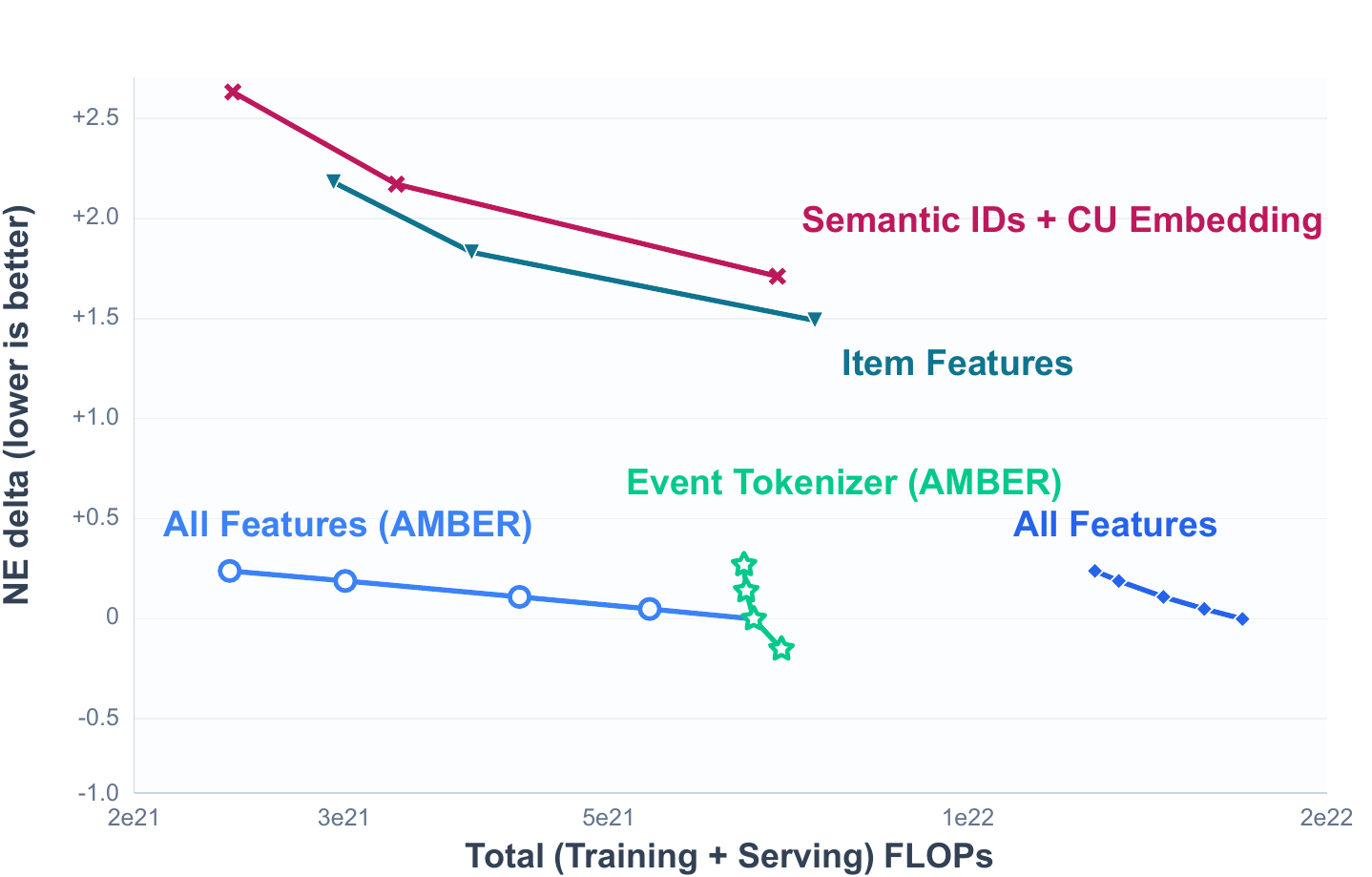}
\Description{Scaling results under total training and serving FLOPs, including CPU feature materialization. More comprehensive feature sets improve NE, while asynchronous Event Token computation and caching advance the Pareto frontier and favor Event Tokenizer scaling over User LLM scaling.}
\caption{Scaling snapshot resolution under total training and serving FLOPs, including CPU feature materialization. Moving from Semantic IDs and content-understanding embeddings to item features and then full event features progressively advances the Pareto frontier. By asynchronously computing and caching Event Tokens, AMBER makes full-feature scaling practical and gives Event Tokenizer scaling a more favorable compute-quality trend than User LLM scaling (Sec.~\ref{subsec:scaling}).}
\label{fig:main_scale_up}
\end{figure}

\section{Introduction}
\label{section:intro}


Large autoregressive models are increasingly applied to sequential prediction tasks that operate over real-world \textit{events}: recommendation model sequences of user interaction events~\citep{hstu, hllm, tiger}, world models predict future states from observation-action sequences~\citep{jepa, dreamerv3, genie}, and financial models forecast market movements from financial event sequences~\citep{stockgpt}. These domains share a common structure: each sequence position encodes a \textbf{temporal snapshot}---a structured record of the system's observations at event time---collectively described by a heterogeneous set of signals. We refer to foundational models that predict future states from structured event sequences as \textbf{Large Event Models (LEMs)}. While LLMs operate over a fixed vocabulary, LEMs operate over temporal snapshots of real-world events.


The distinction between LEMs and LLMs lies in what we call \textbf{snapshot resolution}. At each moment, the true system state contains far more information than any system can record; snapshot resolution is how much of it a model actually captures---formally, the number of distinct signals encoded per event. For example, recommendation events contain user, item, context, and outcome signals, while robotic-control events contain visual observations, joint states, and active subgoals. Under autoregressive training, higher snapshot resolution at each position improves both sides of the attention mechanism: the query expresses more fine-grained intent, and the key-value history retains more contextual detail for all downstream positions.

Industrial recommendation---a representative large-scale LEM instance---currently faces a strict trade-off between sequence length and snapshot resolution, driven by serving compute requirements: in large-scale recommendation systems, feature-related operations (storage, deserialization, and network transfer) can rival or even exceed GPU compute. This bottleneck forces architectures into compromises. Traditional pointwise models~\citep{din, bst} retain comprehensive heterogeneous features for the current query but represent the user history sparsely. Conversely, recent autoregressive models excel at sequence modeling but are forced to reduce detailed historical events to simple item-semantic representations~\citep{tiger, lcrec, onerec, hllm} or manually curated feature subsets~\citep{hstu, plum}. In both cases, this forced information loss weakens individual predictions, and because each position acts as context for the next, the degradation compounds progressively across the sequence.


Prior work demonstrates that LLMs can process compressed latent representations, such as visual tokens, after alignment with their embedding space~\citep{llava, blip2}. Therefore, we propose \textbf{AMBER} (\textbf{A}utoregressive \textbf{M}odeling via \textbf{B}ottlenecked \textbf{E}vent \textbf{R}epresentation), an architecture centered around a unified Event Tokenizer. Rather than feeding raw features directly to the sequence model, the tokenizer encapsulates the vast heterogeneous signals of each temporal snapshot into a compact Event Token, which is then consumed autoregressively by the downstream sequence model. The entire pipeline is trained end-to-end, learning an efficiently compressed event token representation that scales effectively with downstream LLM capacity. At serving time, because tokenization is per-event and history-independent, it is triggered asynchronously as events occur, and the resulting Event Tokens are cached into the user's history sequence (Figure~\ref{fig:amber_flow}). Consequently, caching each Event Token after a single asynchronous computation avoids materializing raw features during real-time inference, decoupling snapshot resolution from serving compute requirements.

We verify that \textbf{snapshot resolution} is a viable and effective scaling dimension for industrial recommendation. On top of this finding, we demonstrate that the event token paradigm has the following additional properties that make it well-suited for industrial recommendation at scale:

\textbf{First, Event Tokenizer capacity can be scaled independently to improve downstream performance with minimal impact on real-time serving compute.} Since tokenization is triggered once per impression and cached, scaling Event Tokenizer capacity yields a favorable compute-quality trade-off. It yields substantial downstream quality improvements while adding negligible compute requirements to the primary online serving path.

\textbf{Second, downstream LLM performance scales predictably with capacity over Event Tokens.} By treating these compact tokens as a standardized continuous input modality, the downstream sequential model yields monotonic performance improvements as its parameter capacity increases. This indicates that rather than experiencing excessive compression, the Event Tokens retain sufficient fine-grained semantic information to successfully unlock the scaling laws of the downstream model.


In addition to these contributions, we share practical lessons from operating AMBER in a large-scale recommendation system, focusing on recurrent representation drift and Event Token storage. Periodic tokenizer retraining required by open vocabularies and evolving user trends shifts the representation space, making newly generated tokens incompatible with cached histories. We mitigate this instability with an adversarial recurrent alignment strategy that constrains drift across encoder updates. We also apply Matryoshka Dropout and quantization-aware training to reduce Event Token storage.


We evaluate AMBER in a large-scale production environment, covering both ranking and retrieval. Under total training and serving compute, AMBER advances the compute-quality Pareto frontier relative to alternative recommendation paradigms. By aligning Event Tokens with the LLM embedding space, AMBER enables pretrained weights to outperform random initialization, supporting Event Tokens as a new LLM input modality. Large-scale evaluation further shows that an existing non-LLM ranker can directly consume Event Tokens as additional historical features, yielding a statistically significant improvement.


In summary, AMBER introduces event-level tokenization to incorporate heterogeneous features into event sequences, enables compute-efficient scaling through asynchronous tokenization, and addresses the representation drift and storage challenges of large-scale use.

\section{Related Work}
\label{section:related}

\paragraph{Sequence Modeling in Recommendation} Industrial recommenders primarily rely on two feature sets: per-impression heterogeneous features and historical behavior features. Due to strict serving-compute constraints, the snapshot resolution of a per-impression query is typically tens to hundreds of times larger than that of a historical event. This asymmetry forces a strict modeling dichotomy. \textbf{Pointwise models}~\citep{din, bst} use the high-resolution per-impression features to attend over the low-resolution history; while this yields strong performance, it severely degrades training efficiency by redundantly re-encoding the history for every impression. Conversely, \textbf{autoregressive models}~\citep{sasrec, hstu} process history sequentially in an efficient single causal pass, but the query snapshot resolution at each position is inherently weak, which ultimately caps overall performance. AMBER bridges this gap, achieving the single-pass training efficiency of autoregressive modeling while unlocking the high per-event resolution characteristic of pointwise models.

\paragraph{LLM Recommendation} Most LLM-based recommenders serialize events into text~\citep{p5} or Semantic IDs~\citep{tiger, onerec, lcrec, plum} or both. However, bounded context windows under strict serving constraints force a direct trade-off between per-event token budget and history coverage. Consequently, both paradigms compromise: Semantic IDs efficiently encode items but discard detailed event context, while text can describe full events but suffers from low information density. Similar to multimodal LLMs~\citep{llava, blip2, flamingo}, HLLM~\citep{hllm} trains end-to-end to compress item text into embeddings but discards relevant event-level signals. Concurrent work, LoopFM~\citep{loopfm}, forms sequences from intermediate foundation model (FM) representations; however, this decoupled distillation blocks downstream gradients, diminishing improvements when the downstream model matches the FM's capacity. AMBER overcomes both by training an \emph{event-centric} tokenizer end-to-end, compressing the full heterogeneous event into high-density Event Tokens optimized directly for downstream sequential modeling.

\paragraph{Representation Drifting and System Efficiency} Caching learned representations introduces two systemic challenges: representation drift during recurrent training and substantial storage requirements at scale. To address temporal drift, we draw on Domain Adaptation~\citep{dann, mmd, coral}, a field that aligns feature spaces across varying distributions. While traditionally applied to cross-domain tasks, we adapt these adversarial principles~\citep{dann} to stabilize representations across successive temporal checkpoints. To resolve the storage bottleneck, we build upon recent advances in Representation Compression, specifically Matryoshka Representation Learning~\citep{matryoshka, matryoshka_rec} and quantization-aware training~\citep{ste}, to substantially reduce the memory footprint of cached tokens.

\section{Methodology}
\label{section:method}

\subsection{AMBER: Model Architecture}
\label{subsec:overview}

Current industrial recommendation systems exhibit a substantial computational asymmetry between training and serving. Because offline training operates without strict latency constraints and processes only actual logged behaviors alongside synthetic negatives, its computational cost is merely a fraction of online serving, which score large candidate pools in real time. Our architecture exploits this structural asymmetry by deliberately using additional offline training compute to reduce online serving compute requirements.

\begin{figure}[t]
\centering
\includegraphics[width=\linewidth]{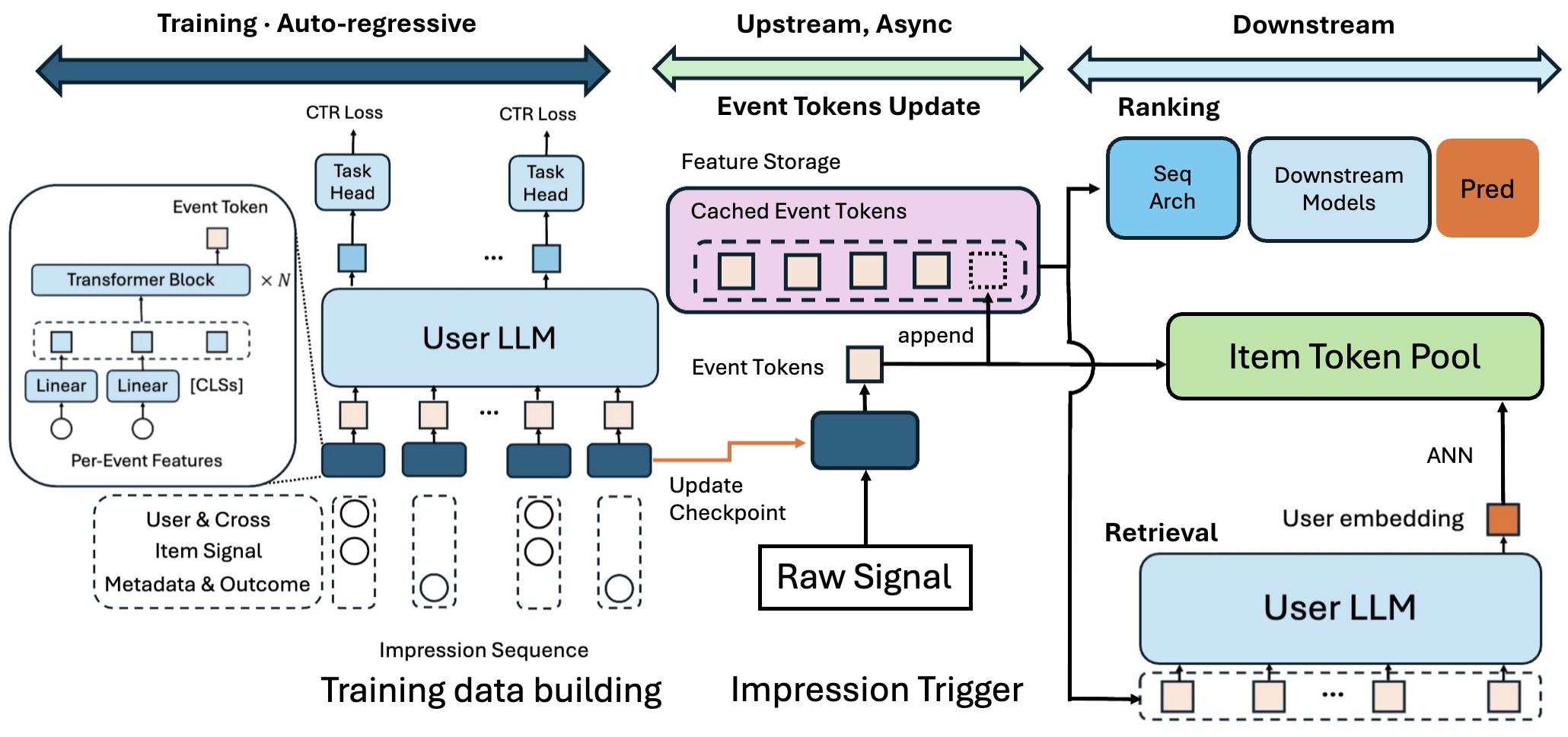}
\Description{AMBER flow diagram spanning training, asynchronous upstream processing, and downstream modeling.}
\caption{Overview of AMBER. \textbf{Training} (left): the Event Tokenizer is trained end-to-end with the User LLM under an autoregressive objective, learning to compress heterogeneous features into latent embeddings (Event Tokens). \textbf{Asynchronous upstream} (middle): when an event occurs, the Event Tokenizer compresses its full feature set into a compact Event Token and appends it to the user's existing Event Token sequence; item tokens are also used to update the existing item-token pool. \textbf{Downstream} (right): ranking and retrieval models consume cached Event Token sequences as sequential features. By compressing full event features into Event Tokens asynchronously, AMBER increases the snapshot resolution of user histories without materializing the full feature set at serving time, thereby avoiding the associated serving compute (Sec.~\ref{subsec:overview}).}
\label{fig:amber_flow}
\end{figure}

Following this philosophy, we propose \textbf{AMBER}, an event-centric paradigm that uses additional offline training to encapsulate each interaction's heterogeneous features into compact tokens, so that at serving time the downstream model consumes cached Event Tokens instead of materializing detailed per-event signals, achieving comparable quality at far lower serving compute. It has two components:
\begin{itemize}
    \item \textbf{Event Tokenizer} $g_\theta$: compresses the raw per-event signals into $d_z$-dimensional \textbf{Event Tokens} $\mathbf{z}_i$.
    \item \textbf{User LLM} $f_\psi$: consumes the Event Tokens and makes predictions autoregressively.
\end{itemize}
In our experiments, $f_\psi$ is initialized from a pre-trained 1B-parameter Llama model. For RoPE, we use event timestamps relative to a fixed reference date rather than sequence indices.
During training, the two models ($g_\theta$, $f_\psi$) are jointly optimized. During serving, they are decoupled: the Event Tokenizer runs asynchronously, and the downstream model makes predictions from the cached Event Token sequence (Figure~\ref{fig:amber_flow}). As the User LLM is a decoder-only Transformer, the rest of this subsection focuses on the Event Tokenizer.

\paragraph{Tokenizer architecture}
The Event Tokenizer maps raw signals to an Event Token in three steps (Figure~\ref{fig:encoder_architectures}(a); detailed in Appendix~\ref{appendix:baselines}):
\begin{enumerate}
    \item \textbf{Feature encoding.} Sparse categorical features (via embedding lookup), embedding features (via linear projection), and dense numerical features (normalized, concatenated, and projected) are each mapped to independent $d_{\text{model}}$-dimensional tokens $\mathbf{h}_1, \ldots, \mathbf{h}_m$.
    \item \textbf{Interaction \& summarizing.} We append $c$ learnable \texttt{[CLS]} tokens $\mathbf{h}_{\texttt{CLS}}^{(1)}, \ldots, \mathbf{h}_{\texttt{CLS}}^{(c)}$ to the feature tokens and process the sequence through an $L$-layer bidirectional Transformer encoder, so features interact and are summarized into the $c$ \texttt{[CLS]} outputs.
    \item \textbf{LLM-space projection.} An MLP maps the contextualized \texttt{[CLS]} outputs to the final $d_z$-dimensional Event Token.
\end{enumerate}
Concretely, the Event Token is
\begin{equation}
    \mathbf{z}_i = \text{MLP}\!\left(\text{BiTransformer}(\mathbf{h}_1, \ldots, \mathbf{h}_m, \mathbf{h}_{\texttt{CLS}}^{(1:c)})[\texttt{CLS}]\right).
    \label{eq:cls}
\end{equation}
We also compare against other architectures in Section~\ref{section:experiments}.

\paragraph{Unified Event Tokenizer}
Instead of task-specific encoders, a single shared Event Tokenizer $g_\theta$ generates tokens via dynamic feature masking:
\begin{equation}
    \mathbf{z}_i = g_\theta(\mathbf{f}_i, \mathbf{m}), \quad \mathbf{z}_i \in \mathbb{R}^{d_z}
    \label{eq:event_token}
\end{equation}
where $\mathbf{f}_i$ denotes the heterogeneous features of event $e_i$, and $\mathbf{m}$ is a role based binary mask dictating feature visibility (Table~\ref{tab:masking}). Beyond simplifying serving, this shared design outperforms dedicated per-entity encoders via positive transfer across entity(\S\ref{subsec:training_design}).

\subsection{Training and Sequence Design}
\label{subsec:ranking}\label{subsec:training}\label{subsubsec:training_stages}

\paragraph{Three-stage training} Directly training a randomly initialized Event Tokenizer with a pre-trained LLM would corrupt the LLM's world knowledge. We therefore use two-stage alignment inspired by vision-language model training~\citep{llava, blip2}, followed by recurrent training:\par

\indent\textbf{Stage~1 (Pre-alignment).} We freeze the User LLM and train only the Event Tokenizer to align event features with the LLM's pre-trained embedding space.

\indent\textbf{Stage~2 (Joint training).} We unfreeze the User LLM and train both components end-to-end to co-adapt their representations.

\indent\textbf{Stage~3 (Recurrent training).} We periodically retrain to track distribution shifts. Section~\ref{section:serving} addresses the resulting drift of cached Event Tokens across updates.

\begin{figure}[t]
\centering
\includegraphics[width=\linewidth]{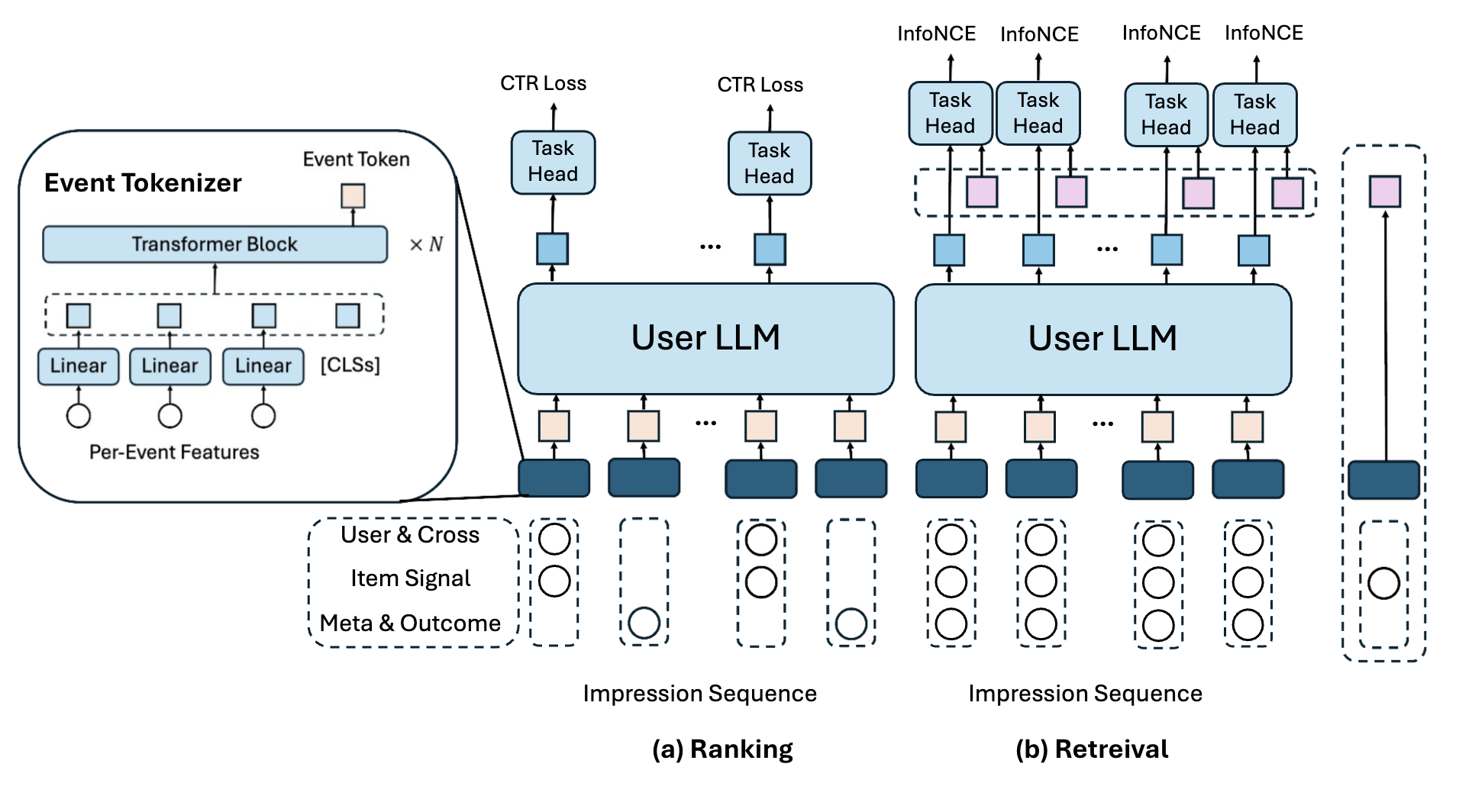}
\Description{AMBER architecture for the two tasks: (a) ranking with an interleaved sequence of context and label tokens trained with BCE, and (b) retrieval with fused event tokens trained with InfoNCE and a separately tokenized candidate item.}
\caption{AMBER for two tasks. (a)~Ranking uses an interleaved sequence of context and label tokens with a BCE objective. (b)~Retrieval uses fused event tokens with an InfoNCE objective; the candidate item (pink) is tokenized separately. All Event Tokenizer instances share weights and differ only in their feature masks for different token types (Sec.~\ref{subsec:ranking}).}
\label{fig:architecture}
\end{figure}

\paragraph{Sequence Design} Context length is a bottleneck in sequential recommendation: under a fixed context window, every extra token per event evicts a historical event. Our experiments (Section~\ref{subsec:training_design}) show that prioritizing historical coverage over tokens per event yields better performance. We seek the minimum token budget each task requires, leading to one token per event for retrieval and two tokens per event for ranking (Figure~\ref{fig:architecture}, Table~\ref{tab:masking}).

\begin{table}[t]
\centering
\small
\setlength{\tabcolsep}{3pt}
\caption{Per-role feature masking. A single shared Event Tokenizer serves all roles; only the input mask differs. \cmark: kept, \xmark: masked (metadata $=$ position, weight) (Sec.~\ref{subsec:ranking}).}
\label{tab:masking}
\begin{tabular}{@{}lcccc@{}}
\toprule
\textbf{Token role} & \textbf{User} & \textbf{Item} & \textbf{Ctx \& Cross} & \textbf{Metadata \& Outcome} \\
\midrule
Ranking context (A) & \cmark & \cmark & \cmark & \xmark \\
Ranking label (B)   & \xmark & \xmark & \xmark & \cmark \\
Retrieval history   & \cmark & \cmark & \cmark & \cmark \\
Retrieval target    & \xmark & \cmark & \xmark & \xmark \\
\bottomrule
\end{tabular}
\end{table}

\paragraph{Ranking objective (autoregressive BCE)}
We interleave, for each event, a context token $\mathbf{z}_i^{(\text{A})}$ and a label token $\mathbf{z}_i^{(\text{B})}$ under complementary masks (Table~\ref{tab:masking}), forming the sequence
\begin{equation}
    (\mathbf{z}_1^{(\text{A})}, \mathbf{z}_1^{(\text{B})}, \mathbf{z}_2^{(\text{A})}, \mathbf{z}_2^{(\text{B})}, \ldots, \mathbf{z}_n^{(\text{A})}, \mathbf{z}_n^{(\text{B})}).
    \label{eq:interleave}
\end{equation}
Each context position $\mathbf{z}_i^{(\text{A})}$ predicts event~$i$'s label via an MLP head. Since the label token $\mathbf{z}_i^{(\text{B})}$ follows $\mathbf{z}_i^{(\text{A})}$ in the sequence, it is never attended to, keeping the prediction leakage-free. Training runs over the whole sequence in a single forward pass:
\begin{equation}
    \mathcal{L}_{\text{rank}} = -\frac{1}{n} \sum_{i=1}^{n} \left[ y_{i} \log \hat{y}_{i} + (1 - y_{i}) \log (1 - \hat{y}_{i}) \right],
    \label{eq:bce}
\end{equation}
where $y_{i} \in \{0, 1\}$ is the behavior label. At inference, the candidate is appended as a context token with no label.

\paragraph{Retrieval objective (InfoNCE)}
We fuse each event into a single unmasked token, forming the sequence
\begin{equation}
    (\mathbf{z}_1, \mathbf{z}_2, \ldots, \mathbf{z}_n),
    \label{eq:fused}
\end{equation}
which maximizes the history the LLM attends over. At each position the LLM emits a user embedding $\mathbf{u}_i$; targets are pre-encoded by the same tokenizer from item-only features and cached independently of the user. We train with a noise-contrastive estimation loss:
\begin{equation}
    \mathcal{L}_{\text{ret}} = -\frac{1}{n} \sum_{i=1}^{n} \log \frac{\exp(\mathbf{u}_i^\top \mathbf{v}^+_{i} / \tau)}{\exp(\mathbf{u}_i^\top \mathbf{v}^+_{i} / \tau) + \sum_{j \in \mathcal{N}_i} \exp(\mathbf{u}_i^\top \mathbf{v}^-_{j} / \tau)},
    \label{eq:nce}
\end{equation}
where $\mathbf{v}^+_i$ is a positively engaged item in the user's near future, $\tau$ is a temperature, and $\mathcal{N}_i$ combines \textit{hard negatives} (items shown but not engaged) and \textit{easy negatives} (randomly sampled items).

\section{Serving at Scale}
\label{section:serving}

\subsection{Asynchronous Upstream Architecture}
\label{subsec:system}

AMBER operates as a two-stage system that decouples feature computation from real-time serving:\par

\indent\textbf{Offline tokenization and caching.} The Event Tokenizer runs asynchronously: triggered when an impression event occurs, it queries the event's full heterogeneous feature set through the existing training-data pipeline and compresses it into an Event Token. Each token is appended to the user's Event Token sequence in the feature store.

\indent\textbf{Downstream modes.} Cached Event Tokens support two downstream modes:\par
\begin{itemize}
    \item \textit{Ranking.} Serving end-to-end LLMs for real-time ranking introduces substantial infrastructure bottlenecks, particularly regarding KV cache maintenance for long user histories. While resolving these serving challenges is beyond the scope of this work, we demonstrate the universal transferability of our representation: cached Event Tokens are served as plug-and-play sequential features to the existing non-LLM ranking model (validated in \S\ref{subsec:online}).
    \item \textit{Retrieval.} The User LLM directly consumes the cached token sequence to produce real-time user embeddings for approximate nearest-neighbor (ANN) search.
\end{itemize}

\subsection{Mitigating Representation Drift}
\label{subsec:drift}

\begin{figure}[t]
\centering
\includegraphics[width=\linewidth]{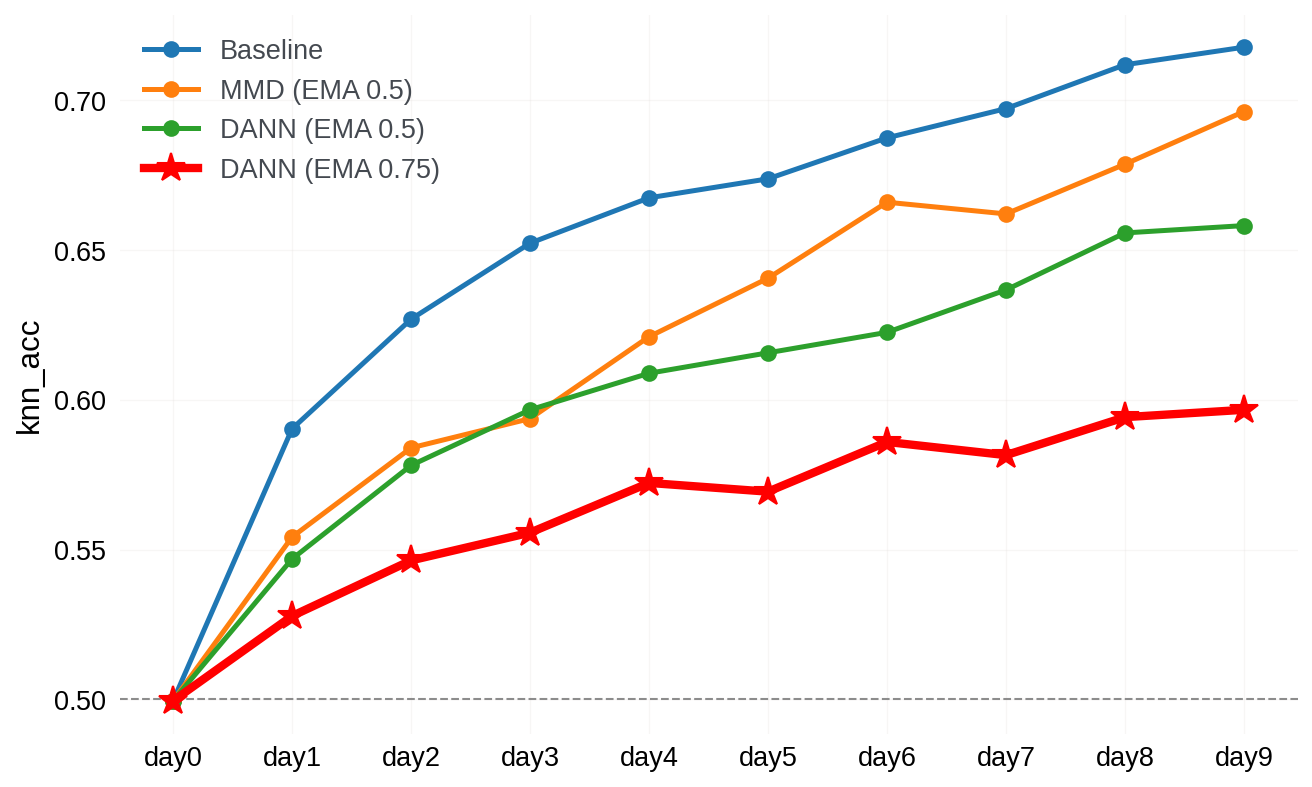}
\Description{Line plot of checkpoint k-NN accuracy over ten recurrent-training days. The baseline exhibits the greatest drift, while MMD and DANN with EMA reduce checkpoint separability; DANN with EMA 0.75 remains closest to 0.5.}
\caption{Mitigating representation drift over time, measured by checkpoint separability ($k$-NN accuracy; higher $=$ more drift, $0.5$ $=$ indistinguishable). The unconstrained baseline (no regularization) rises monotonically, indicating severe progressive drift in the learned embedding space. Our stabilization configurations (EMA($0.5$)$+$DANN, EMA($0.5$)$+$MMD, and EMA($0.75$)$+$DANN) suppress this separation, keeping the representation consistent across continuous retraining cycles (Sec.~\ref{subsec:drift}).}
\label{fig:drift_curve}
\end{figure}

During recurrent training, updating the encoder from $g_{\theta_{\text{old}}}$ to $g_{\theta_{\text{new}}}$ alters its output space, causing \emph{representation drift} that accumulates across retraining (Figure~\ref{fig:drift_curve}) and renders tokenized events incompatible with previously cached tokens (t-SNE visualization in Appendix~\ref{appendix:tsne}), destabilizing downstream LLM predictions (Appendix~\ref{appendix:stability}).

To mitigate representation drift, we combine \textbf{adversarial regularization} based on domain-adversarial neural network (DANN) training~\citep{dann} with \textbf{exponential moving average (EMA) updates}.

\paragraph{Adversarial regularization} During recurrent training, a lightweight discriminator $D_\phi$ is trained to distinguish whether a token was produced by the old or new encoder:
\begin{equation}
    \mathcal{L}_{\text{adv}} = -\frac{1}{n} \sum_{i=1}^{n} \left[ s_i \log D_\phi(\mathbf{z}_i) + (1 - s_i) \log (1 - D_\phi(\mathbf{z}_i)) \right],
    \label{eq:adv}
\end{equation}
where $s_i \in \{0, 1\}$ indicates the encoder version. A gradient reversal layer pushes $g_{\theta_{\text{new}}}$ to produce representations indistinguishable from $g_{\theta_{\text{old}}}$, minimizing the joint objective $\mathcal{L}_{\text{task}} + \lambda\,\mathcal{L}_{\text{adv}}$. We use a small $\lambda$ (${\sim}5\times10^{-3}$), as larger values can produce degenerate representations that fool the discriminator without reducing drift.

\paragraph{EMA weight update} We maintain an EMA encoder across daily cycles as the stable reference to which DANN aligns each new encoder, rather than anchoring only to the previous day.

We evaluate DANN against alternative drift-mitigation objectives in Section~\ref{subsec:drift_evaluation}.

\paragraph{Sharded recurrent training} Retraining on the full user base every day is wasteful: only a small fraction of each user's sequence is new, so most data is re-seen. We instead randomly partition users into $N$ shards and retrain on one shard per day, cycling through all users every $N$ days to maximizing the fresh signal per training. When model converged, we find this suffices to catch up to the latest trends.

\subsection{Embedding Compression}
\label{subsec:compression}

Storing hundreds of pre-computed Event Tokens per user requires compression to improve storage efficiency. We apply two complementary techniques.

\paragraph{Matryoshka Dropout} We introduce Matryoshka Dropout to support flexible token dimensions without repeatedly retraining the model. Standard Matryoshka Learning~\citep{matryoshka} requires multiple downstream forward passes per step, a computationally prohibitive operation for LLMs. We instead achieve continuous ordering in a single pass via structured suffix dropout. With probability $\rho$, we truncate $\mathbf{z}$ to its leading $d_r \sim \text{Uniform}(d_{\min}, d_z)$ dimensions and apply $d_z/d_r$ rescaling:
\begin{equation}
    \tilde{\mathbf{z}}^{(j)} = \begin{cases}
        \frac{d_z}{d_r}\,\mathbf{z}^{(j)} & \text{if } j \leq d_r, \\
        0 & \text{if } j > d_r,
    \end{cases}
    \label{eq:matryoshka}
\end{equation}
so the model learns to retain the most important information in each prefix.

\paragraph{Quantization-aware training} We expose the encoder to simulated INT8 noise via STE~\citep{ste} during training and serve at INT4, achieving an $8\times$ storage reduction (vs.\ FP32) while preserving ${\sim}80\%$ of the full-precision embedding's impact.


\section{Experiments}
\label{section:experiments}

We evaluate AMBER via a \textbf{dual-track} validation:\par

\indent\textbf{Model Effectiveness and Scaling} (\S\ref{subsec:main_results}--\S\ref{subsec:scaling}). We benchmark AMBER against a highly optimized ranking model (hereafter, \emph{Incumbent}) and other paradigms (\S\ref{subsec:main_results}), conduct ablation studies of input features and tokenizer designs (\S\ref{subsec:analysis}--\S\ref{subsec:training_design}), and analyze scaling behavior across tokenizer capacity, feature count, and downstream model size (\S\ref{subsec:scaling}).

\indent\textbf{Cross-Architecture Transfer} (\S\ref{subsec:online}). We demonstrate the model-agnostic nature of Event Tokens by serving the decoupled tokenizer as plug-and-play features for an existing non-LLM ranker.

\subsection{Setup}
\label{subsec:setup}

\subsubsection{Dataset}
Models are trained on petabyte-scale impression logs from a large-scale industrial recommendation platform using chronological splits: training and feature selection use past data, calibration uses only recent data preceding evaluation, and all metrics are computed on unseen future dates. Each event carries the few hundred most important features across user, item, context, and outcome signals; ablations (\S\ref{subsec:analysis}--\S\ref{subsec:training_design}) use a 10\% subsample, while main results (\S\ref{subsec:main_results}) use the full dataset.

AMBER targets real-world system constraints that arise from feature materialization, recurrent training, and non-stationarity. Evaluating these constraints requires hundreds of heterogeneous features per event and billion-scale temporal data. Petabyte-scale industrial data provides both properties, which common public recommendation benchmarks do not jointly expose.

\subsubsection{Metrics}
For ranking, our primary evaluation metric is \textbf{Normalized Entropy (NE)}~\citep{he2014practical}, a standard offline metric widely used as a proxy for online ranking quality in industrial recommendation. NE is defined as the average log loss normalized by the entropy of the empirical positive rate $p$:
\begin{equation}
    \text{NE} = \frac{-\frac{1}{N} \sum_{i=1}^{N} \left[ y_i \log \hat{y}_i + (1 - y_i) \log(1 - \hat{y}_i) \right]}{-\left[ p \log p + (1 - p) \log(1 - p) \right]},
    \label{eq:ne}
\end{equation}
where $y_i \in \{0, 1\}$ is the ground-truth label, and $p$ is the empirical interaction rate observed in the evaluation dataset. All NE improvements are reported as relative deltas ($\text{NE }\Delta$). We use the following metrics:\par
\indent\textbf{NE.} Because NE is computed from log loss, it reflects both ranking quality and probability calibration: over- or under-confident predictions increase NE even when the ranking remains unchanged. We therefore calibrate each model on leakage-free data from the recent past so that the aggregate predicted interaction rate matches the observed interaction rate (i.e., a calibration ratio close to 1), and compute NE on unseen future data.

\indent\textbf{Ensemble NE.} Direct cross-paradigm comparison is insufficient to assess complementary predictive benefit. Therefore, Ensemble NE measures the complementary predictive benefit of AMBER relative to the highly optimized Incumbent. On unseen future data, we compute NE after linearly blending their predictions, with the mixing weight tuned on a preceding held-out slice. To account for feature-materialization latency, we mask preceding events within either a \textbf{10-minute} or \textbf{1-minute} delay window. Lower Ensemble NE indicates that AMBER captures additional signal over the Incumbent. We leverage this as an offline measure of complementary predictive benefit.

\indent\textbf{Soft Recall.} For retrieval, we use an all-candidate evaluation set in which every candidate receives its counterfactual final-stage auction value. Within each request group, Soft Recall is the fraction of nonnegative value from the ideal auction top-$K$ retained in the retriever's top-$N$. It measures potential auction value preserved through retrieval rather than ID overlap.

NE differences within $0.02\%$ are averaged over two runs; the observed run-to-run variation is at most $0.02\%$.

\subsubsection{Baselines}\leavevmode\par
\indent\textbf{Incumbent models.} \emph{Incumbent} denotes the highly optimized ranking model, trained on trillions of examples, against which we report Ensemble NE. \emph{Incumbent (fair)} matches AMBER on features, FLOP budget, and training target, serving as the controlled baseline for NE.

\indent\textbf{HSTU-style input.} This baseline follows HSTU's input schema (item categorical features per event and user features once at the sequence front) while retaining AMBER's LLM to isolate the input schema; prior work finds Transformer and HSTU backbones comparable at scale~\citep{argus} (Appendix~\ref{appendix:hstulike})\footnote{The downstream sequence-model architecture is not a focus of AMBER. The Event Tokenizer can also be paired with HSTU as the downstream sequence model. For a controlled comparison, we follow the existing HSTU input schema while using the same 1B-parameter LLM as AMBER.}.

\indent\textbf{Semantic IDs + CU embeddings.} Our ablation study (\S\ref{subsec:training_design}) shows that history coverage contributes more than additional tokens per event under a fixed context length. We therefore fuse multiple Semantic IDs and content-understanding (CU) embeddings into a single token. CU embeddings are included as a lossless upper bound for Semantic ID decompression. To control for tokenizer architecture and capacity, this baseline uses the same Event Tokenizer architecture and scale as AMBER.

\indent\textbf{Pointwise}$^\dagger$\textbf{.} To separate autoregressive training efficiency from history resolution, this variant reproduces the pointwise input pattern within AMBER's backbone: the current query uses the full event, while history retains only item features. We include the complete item-feature set as a strong upper bound, although materializing this many historical features online would be computationally costly (Appendix~\ref{appendix:pointwise}).

\subsection{Main Results}
\label{subsec:main_results}

\subsubsection{Ranking}

As established in \S\ref{subsec:setup}, we evaluate both controlled model quality and complementary predictive signal. Table~\ref{tab:main} shows that AMBER improves on both fronts.

\begin{table}[t]
\centering
\caption{Ranking performance. NE $\Delta$ is reported relative to Incumbent (fair). Ensemble NE $\Delta$ measures the complementary predictive signal provided when linearly blending predictions with the Incumbent, evaluated under simulated 10-minute and 1-minute feature delays (Sec.~\ref{subsec:main_results}).}
\label{tab:main}
\small
\setlength{\tabcolsep}{4pt}
\begin{tabular}{lccc}
\toprule
& \textbf{NE} $\Delta\,\downarrow$ & \multicolumn{2}{c}{\textbf{Ensemble NE} $\Delta\,\downarrow$ (vs.\ Incumbent)} \\
\cmidrule(lr){2-2} \cmidrule(lr){3-4}
\textbf{Method} & \footnotesize vs.\ Incumbent (fair) & \footnotesize 10-min delay & \footnotesize 1-min delay \\
\midrule
SID $+$ CU emb & $+1.20\%$ & $-0.03\%$ & $-0.05\%$ \\
HSTU-style & $+0.60\%$ & $-0.03\%$ & $-0.06\%$ \\
Pointwise$^\dagger$ & $-0.10\%$ & --- & --- \\
AMBER & $\mathbf{-0.40\%}$ & $\mathbf{-0.10\%}$ & $\mathbf{-0.16\%}$ \\
\bottomrule
\end{tabular}
\begin{minipage}{\linewidth}
\footnotesize\raggedright
$^\dagger$ See Appendix~\ref{appendix:pointwise} for the Pointwise setup.
\end{minipage}
\end{table}

\paragraph{Performance against Controlled Baselines} AMBER achieves the lowest NE, outperforming Incumbent (fair), Pointwise, HSTU-style input, and Semantic IDs + CU embeddings by $0.40\%$, $0.30\%$, $1.00\%$, and $1.60\%$, respectively. The $1.60\%$ improvement over Semantic IDs + CU embeddings highlights the information bottleneck of item-centric representations. AMBER's $0.30\%$ edge over Pointwise confirms that maintaining high snapshot resolution throughout the history---rather than solely at the current query---is beneficial. Meanwhile, Pointwise's $0.10\%$ improvement over Incumbent (fair) isolates the efficiency benefit of autoregressive modeling under matched compute. Finally, while HSTU-style input reduces NE by $0.60\%$ over Semantic IDs + CU embeddings through static user-side and item categorical features, its similar Ensemble NE suggests that these signals are largely redundant with the Incumbent.

\paragraph{Complementary Signal to the Incumbent}
Blending AMBER with the Incumbent reduces Ensemble NE by $0.10\%$ under a 10-minute feature delay and $0.16\%$ under a 1-minute window. In contrast, Semantic IDs + CU embeddings yield only $0.03\%$ and $0.05\%$ reductions—less than one-third of AMBER's improvements—and degrade further without CU embeddings (\S\ref{subsubsec:featureablation}). Despite this gap, the item-semantic baseline still provides non-zero ensemble improvements, which may partly result from the LLM capacity.

\subsubsection{Retrieval}

We evaluate retrieval via Soft Recall (\S\ref{subsec:setup}). AMBER maintains a $0.31$--$0.51\%$ improvement across all checkpoint ages, outperforming CU embeddings by $0.28$--$0.32$ percentage points (Table~\ref{tab:retrieval}). While the latter becomes neutral by Day~8, AMBER retains a $0.31\%$ improvement, indicating that high-resolution event information improves retrieval and remains robust to checkpoint staleness.

\begin{table}[t]
\centering
\caption{Retrieval Soft Recall $\Delta$ over days since the last retraining, relative to the Incumbent (Sec.~\ref{subsec:main_results}).}
\label{tab:retrieval}
\begin{tabular}{lcccc}
\toprule
\textbf{Soft Recall $\uparrow$} & \textbf{Day 1} & \textbf{Day 2} & \textbf{Day 4} & \textbf{Day 8} \\
\midrule
CU emb. & $+0.23\%$ & $+0.09\%$ & $+0.20\%$ & $-0.01\%$ \\
AMBER & $\mathbf{+0.51\%}$ & $\mathbf{+0.35\%}$ & $\mathbf{+0.50\%}$ & $\mathbf{+0.31\%}$ \\
\bottomrule
\end{tabular}
\end{table}

\subsection{Ablation Study}
\label{subsec:analysis}

We further ablate feature groups and tokenizer architectures to derive AMBER's design choices, followed by a scaling analysis that accounts for both training and serving compute (\S\ref{subsec:scaling}).

\subsubsection{Feature Group Ablation}
\label{subsubsec:featureablation}

We conduct two complementary analyses: (1)~removing one feature group at a time from the full AMBER, and (2)~starting from a minimal Semantic ID representation and progressively adding features. All experiments use the same encoder and pre-alignment setup.

\begin{table}[t]
\centering
\caption{Feature group ablation (pre-alignment). SID denotes Semantic IDs; Ensemble NE is measured against the Incumbent under a 10-minute delay. Leave-one-out results are grouped by feature type, and cumulative addition starts from SID $+$ outcome signals (Sec.~\ref{subsubsec:featureablation}).}
\label{tab:feature}
\small
\setlength{\tabcolsep}{3pt}
\begin{tabular}{lcc}
\toprule
& \textbf{NE} $\Delta\,\downarrow$ & \textbf{Ensemble NE} $\Delta\,\downarrow$ \\
\cmidrule(lr){2-2} \cmidrule(lr){3-3}
\textbf{Configuration} & \footnotesize vs.\ Full AMBER & \footnotesize 10-min delay \\
\midrule
\multicolumn{3}{l}{\emph{Item representation}} \\
$-$ Item features & $+6.65\%$ & -- \\
$-$ Item categorical signals & $+1.00\%$ & -- \\
$-$ Item emb & $+0.06\%$ & -- \\
$-$ SID & neutral & -- \\
$-$ (SID $+$ item emb.) & $+0.15\%$ & -- \\
$-$ (SID $+$ Content Hashes) & $+0.07\%$ & -- \\
\addlinespace[2pt]
\multicolumn{3}{l}{\emph{Event context}} \\
$-$ User features & $+0.30\%$ & -- \\
$-$ Outcome signals & $+0.24\%$ & -- \\
$-$ Metadata (weight, position) & $+0.08\%$ & -- \\
$-$ User-item cross features & $+0.03\%$ & -- \\
\midrule
\multicolumn{3}{l}{\emph{Cumulative addition}} \\
SID $+$ outcome signals & $+2.40\%$ & $-0.024\%$ \\
\quad $+$ CU emb & $+1.62\%$ & $-0.034\%$ \\
\quad $+$ sparse features & $+1.52\%$ & $-0.036\%$ \\
\quad $+$ item emb & $+1.50\%$ & $-0.036\%$ \\
\quad $+$ user signals & $+0.40\%$ & $-0.060\%$ \\
\bottomrule
\end{tabular}
\end{table}

\paragraph{Item information is important, but Semantic IDs are insufficient} Removing all item features degrades NE by $6.65\%$, confirming that item representation remains the primary signal in sequential recommendation. However, retaining only Semantic IDs and outcome signals incurs a $2.40\%$ gap to full AMBER. Adding CU embeddings and non-CU item categorical features narrows this gap to $1.52\%$, showing that substantial item information is not captured by Semantic IDs alone.

\paragraph{Learned item representations contain largely overlapping signals} Adding a behavior-based item embedding on top of CU embeddings and item categorical features improves NE by only $0.02\%$ ($1.52\% \to 1.50\%$). Likewise, removing Semantic IDs from full AMBER is neutral, but removing them together with item embeddings incurs a $0.15\%$ loss, compared with $0.06\%$ from removing item embeddings alone. This suggests that Semantic IDs provide little information beyond that already encoded by learned item embeddings.

\paragraph{User features capture signals beyond the behavior sequence} Adding user features improves NE by $1.10\%$, showing that user-side state contains information absent from item interactions. Ensemble NE echoes this result: its largest improvements come from user signals and CU embeddings, both computationally costly to materialize online, while behavior-based item embeddings provide little additional information when reused in the event history. The $0.07\%$ loss from jointly removing Semantic IDs and content hashes further shows that identifier types provide complementary information.

\subsubsection{Tokenizer \& Architecture Design}
\label{subsec:training_design}

We validate AMBER's architecture through four data-driven ablations.

\begin{table*}[t]
\centering
\caption{Ablations of AMBER's design. (a)~Contribution of each training stage. (b)~Transfer from sharing one tokenizer across entity types. (c)~Trade-off between per-event token capacity and history coverage. (d)~Comparison of event encoders (Sec.~\ref{subsec:training_design}).}
\label{tab:design_ablation}
\setlength{\tabcolsep}{2.5pt}
\begin{minipage}[t]{0.242\textwidth}
\centering
\scriptsize
\textbf{(a) Training}\\[2pt]
\begin{tabular*}{\linewidth}{@{\extracolsep{\fill}}lc@{}}
\toprule
\textbf{Setting} & \textbf{NE $\Delta\,\downarrow$} \\
\midrule
Tokenizer only vs.\ frozen LLM & $+0.83\%$ \\
Unfrozen vs.\ frozen LLM & $-1.04\%$ \\
Pre-trained vs.\ random init & $-0.20\%$ \\
Pre-aligned vs.\ direct joint & $-0.10\%$ \\
\bottomrule
\end{tabular*}
\end{minipage}\hspace{0.01\textwidth}%
\begin{minipage}[t]{0.242\textwidth}
\centering
\scriptsize
\textbf{(b) Unified vs. Per-Entity}\\[2pt]
\begin{tabular*}{\linewidth}{@{\extracolsep{\fill}}lc@{}}
\toprule
\textbf{Schema} & \textbf{NE $\Delta\,\downarrow$} \\
\midrule
\shortstack[l]{User / item / other / label\\(4 tokens/event)} & $-0.02\%$ \\
\shortstack[l]{Context / label\\(2 tokens/event)} & $-0.02\%$ \\
\bottomrule
\end{tabular*}
\end{minipage}\hspace{0.01\textwidth}%
\begin{minipage}[t]{0.242\textwidth}
\centering
\scriptsize
\textbf{(c) Number of Tokens per Event}\\[2pt]
\begin{tabular*}{\linewidth}{@{\extracolsep{\fill}}lc@{}}
\toprule
\textbf{Context} & \shortstack{\textbf{NE $\Delta\,\downarrow$}\\\textbf{(2 vs. 1 token/event)}} \\
\midrule
Unlimited context & $-0.10\%$ \\
Limited context & $+0.16\%$ \\
\bottomrule
\end{tabular*}
\end{minipage}\hspace{0.01\textwidth}%
\begin{minipage}[t]{0.242\textwidth}
\centering
\scriptsize
\textbf{(d) Encoder}\\[2pt]
\begin{tabular*}{\linewidth}{@{\extracolsep{\fill}}lc@{}}
\toprule
\textbf{Encoder} & \textbf{NE $\Delta\,\downarrow$} \\
\midrule
Bi-Transformer ($\times 4$, baseline) & --- \\
Bi-Transformer ($\times 2$) & $+0.09\%$ \\
Bi-Transformer ($\times 1$) & $+0.19\%$ \\
DHEN ($\times 12$)~\citep{dhen} & $0.00\%$ \\
PMA ($\times 4$)~\citep{set_transformer} & $+0.20\%$ \\
Concat+FFN & $+0.22\%$ \\
\bottomrule
\end{tabular*}
\end{minipage}
\end{table*}

\paragraph{Unified tokenization enables positive transfer} A shared tokenizer improves NE by $0.02\%$ under both the four-token (user, item, other, label) and two-token (context, label) schemas, showing positive transfer across entity types.

\paragraph{History coverage determines the token budget} Two tokens per event help with unlimited context ($0.10\%$) but hurt under a limited window ($0.16\%$); AMBER therefore prioritizes history coverage.

\paragraph{The Transformer provides the best practical trade-off} A four-layer Transformer matches DHEN ($\times 12$) at similar FLOPs while offering greater flexibility and FlashAttention efficiency; PMA and Concat+FFN (Appendix~\ref{appendix:baselines}) perform worse.

\paragraph{Pre-trained knowledge transfers to event modeling} Even when frozen, the LLM improves over the Event Tokenizer alone by $0.83\%$, demonstrating the potential of using Event Tokens directly as LLM inputs. Unfreezing the LLM adds another $1.04\%$. Although the advantage of pre-trained initialization narrows with more data, it stabilizes at $0.20\%$ over random initialization (Appendix~\ref{appendix:pretrain}).

\subsection{Efficiency \& Scaling Analysis}
\label{subsec:scaling}

\paragraph{Scaling protocol} We scale User LLM depth over $\{1,2,4,8,16\}$, Event Tokenizer depth over $\{1,2,4\}$, width over $\{64,128,256,512\}$, and snapshot resolution over $\{100,200,400\}$ features. Features are ranked by permutation importance, measured by the NE increase after shuffling each feature across examples. We preserve the feature-type distribution at every resolution and always retain label and metadata features.

\paragraph{Computational-cost model} We define total computational cost as
\begin{equation}
    C_{\mathrm{total}} = m_u C_{\mathrm{user}} + m_e C_{\mathrm{event}} + C_{\mathrm{mat}},
    \label{eq:total_cost}
\end{equation}
where ranking workloads yield a $30$--$100\times$ serving-to-training ratio; we use the lower bound. Cached tokenization gives $m_e\ll m_u$, and $C_{\mathrm{mat}}$ captures sequence materialization. NE is measured, whereas serving compute is estimated (Appendix~\ref{appendix:cost_model}).

\paragraph{Snapshot resolution improves the compute-quality frontier} Moving from Semantic IDs and CU embeddings to item features and then full features consistently improves NE (Figure~\ref{fig:main_scale_up}). This result indicate model scaling cannot recover information absent from the input representation. AMBER further improves the frontier by moving full-feature materialization off the online serving path.

\begin{figure}[t]
\centering
\includegraphics[width=\linewidth]{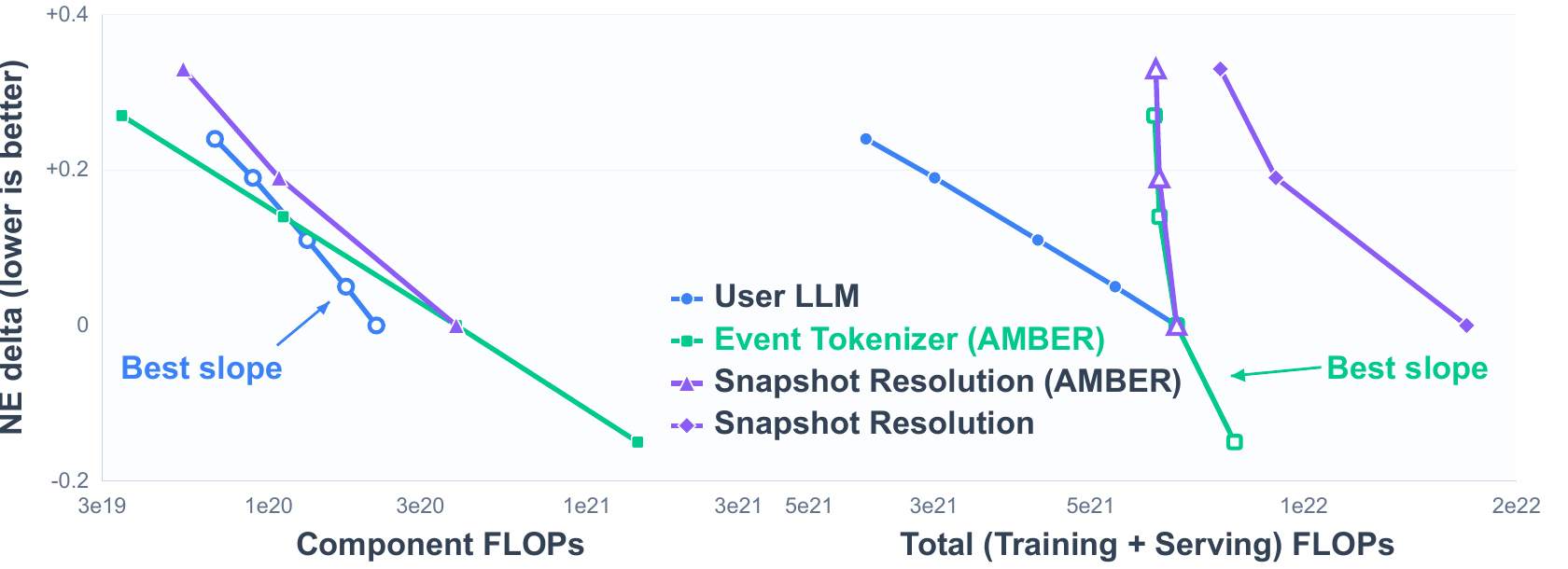}
\Description{Measured User LLM, Event Tokenizer, and snapshot-resolution scaling results under training-only and total-compute accounting. Training FLOPs favor User LLM scaling, whereas total compute favors event-side scaling and AMBER-enabled snapshot resolution.}
\caption{Scaling under training-only and total-compute accounting. User LLM scaling is favored when only training FLOPs are counted (left). Once serving FLOPs are included (right), Event Tokenizer and AMBER-enabled snapshot-resolution scaling become more compute-efficient because tokenization is asynchronous and its outputs are cached (Sec.~\ref{subsec:scaling}).}
\label{fig:train_serving}
\end{figure}

\paragraph{Training-optimal scaling is not system-optimal} When training FLOPs are considered, increasing User LLM capacity provides the strongest scaling trend (Figure~\ref{fig:train_serving}, left). After serving compute is included, event-side scaling becomes more efficient because the User LLM runs for every request, whereas Event Tokens are computed once and cached (Figure~\ref{fig:train_serving}, right). Scaling decisions based only on training compute can therefore be suboptimal under total compute.

\begin{figure}[t]
\centering
\includegraphics[width=\linewidth]{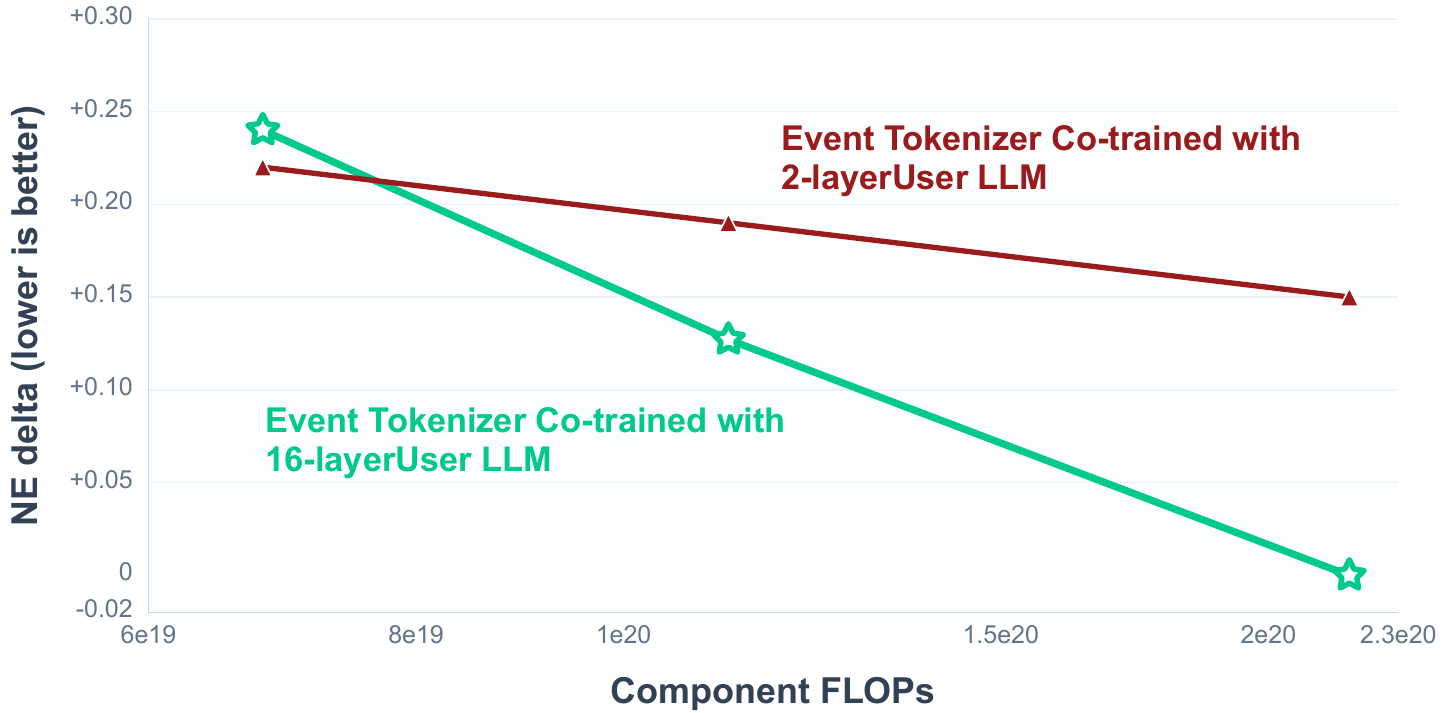}
\Description{Measured downstream User LLM scaling results with frozen Event Tokens co-trained with either a 2-layer or 16-layer LLM. Tokens co-trained with the 16-layer LLM improve more rapidly as downstream compute increases.}
\caption{Scaling downstream User LLMs with frozen Event Tokens. We tokenize future data using Event Tokenizers co-trained with either a 2-layer or 16-layer User LLM, freeze the resulting tokens, and scale a downstream User LLM. Tokens co-trained with the 16-layer model exhibit a stronger scaling trend (Sec.~\ref{subsec:scaling}).}
\label{fig:et_userll_scaling}
\end{figure}

\paragraph{Event Token quality governs downstream scaling} We freeze Event Tokens produced by tokenizers co-trained with either a 2-layer or 16-layer User LLM, and scale a new downstream User LLM on future data. Tokens co-trained with the 16-layer model improve more rapidly with downstream compute (Figure~\ref{fig:et_userll_scaling}), showing that stronger co-training produces representations that larger downstream models can exploit more effectively.

\subsection{Drift Mitigation Evaluation}
\label{subsec:drift_evaluation}

We compare DANN with MMD, CORAL, and $L_2$ regularization over a 20-day checkpoint gap, a stricter and less short-term-sensitive drift test. Cosine similarity compares the same events across encoders; k-NN accuracy independently measures how well a nonlinear classifier separates their representations, reducing dependence on the optimized alignment objective.

\begin{table}[t]
\centering
\caption{Drift mitigation over a 20-day checkpoint gap. Higher cos\_mean and lower knn\_acc indicate better alignment; NE $\Delta$ is relative to no regularization (Sec.~\ref{subsec:drift_evaluation}).}
\label{tab:drift}
\begin{tabular}{lccc}
\toprule
\textbf{Method} & \textbf{cos\_mean $\uparrow$} & \textbf{knn\_acc $\downarrow$} & \textbf{NE $\Delta$ (\%) vs.\ no reg.\ $\downarrow$} \\
\midrule
DANN & \textbf{0.956} & \textbf{0.856} & $\mathbf{-0.02}$ \\
MMD & 0.955 & 0.870 & $+0.03$ \\
CORAL & 0.932 & 0.899 & $+0.36$ \\
$L_2$ & 0.900 & 0.939 & $+1.64$ \\
\bottomrule
\end{tabular}
\end{table}

DANN provides the best alignment without sacrificing predictive quality, and its slight NE improvement may reflect reduced overfitting to the current distribution.

\subsection{Large-Scale Evaluation}
\label{subsec:online}

AMBER's Event Tokenizer (3.84M dense parameters; ${\sim}200$ GB of sparse embedding tables) is deployed on a full-traffic Facebook monetization surface, encoding billions of events online each day. We update it through AMBER's recurrent-training framework and tokenize each logged event once to maintain Event Token sequences. The Incumbent consumes these sequences as additional historical features. AMBER's Event Tokens reduce the Incumbent ranker's NE by approximately $0.06\%$, where $0.02\%$ is considered statistically significant and sufficient to yield measurable conversion lift. This improvement demonstrates cross-architecture transfer and indicates that increasing history snapshot resolution provides predictive signal absent from the Incumbent.


\section{Conclusion}
\label{section:conclusions}

We show that increasing historical event snapshot resolution improves the compute-quality Pareto frontier, but increases high feature-materialization compute during serving. To overcome this, AMBER asynchronously compresses heterogeneous event features into compact Event Tokens. By caching these representations, AMBER enables high-resolution event histories without online feature materialization.

Specifically, AMBER provides three primary advantages: (1)~end-to-end joint training with the LLM enables efficient downstream scaling; (2)~scaling both the LLM and Event Tokenizer reveals that optimizing total training and serving FLOPs, rather than component-level training FLOPs, is important for system-level scaling decisions; and (3)~domain-adversarial training combined with EMA updates mitigates representation drift during recurrent serving. Separately, large-scale evaluation yields a $0.06\%$ NE improvement.

\paragraph{Future Work}
We show that downstream performance scales with Event Tokenizer capacity; future work will explore Mixture-of-Experts (MoE) for more efficient scaling. On the serving side, asynchronous tokenization introduces some feature delay (Appendix~\ref{appendix:delay}). A promising direction is to maintain a high-resolution sequence with very short retention (e.g., 10 minutes), enabling on-demand tokenization of events not yet cached. Finally, because Event Tokens are aligned with the LLM embedding space, they may be jointly modeled with text, effectively treating the user as a new LLM modality. We plan to explore this direction toward personalized LLMs that combine language understanding with recommendation signals.

\bibliographystyle{ACM-Reference-Format}
\bibliography{paper}

\appendix

\section{Total Computational Cost and Sensitivity Analysis}
\label{appendix:cost_model}

We model a ranking system over a common evaluation period. Training uses
logged impressions, heavily samples negatives, and retains only a small
fraction of consideration events. Ranking inference instead runs on every
request without downsampling and scores every candidate passed to the ranking
stage. This gives the workload decomposition
\begin{equation}
    m_u = 1 + r_{\mathrm{traffic}}
        \frac{K_{\mathrm{serve}}}{K_{\mathrm{train}}}
        r_{\mathrm{eff}},
    \label{eq:serving_multiplier}
\end{equation}
where $r_{\mathrm{traffic}}$ is the ratio of served to retained training
impressions, $K_{\mathrm{serve}}/K_{\mathrm{train}}$ captures full-candidate
scoring versus sampled training examples, and $r_{\mathrm{eff}}$ accounts
for the lower utilization of latency-constrained inference relative to batched
training. Empirical workload analysis places $m_u$ on the order of
$30$--$100\times$; we use the lower end in all primary plots. This range is
ranking-specific and is not presented as an industry-wide constant; for
retrieval workloads, this amplification is generally more limited.

The plots compute $C_{\mathrm{user}}$ and $C_{\mathrm{event}}$ from active
parameter FLOPs over the full training run; sparse embedding gathers are
excluded from FLOPs. For AMBER, the Event Tokenizer is invoked asynchronously
and its outputs are cached, so $m_e\ll m_u$. An uncached full-feature sequence
instead sets $m_e=m_u$, reflecting repeated event-side computation on the
ranking path.

Feature-materialization compute depends on the serving stack. We map it to the
common FLOP axis using an empirical power ratio between feature
serving and GPU serving. Because this ratio is infrastructure-dependent, we
report only coarse ranges rather than absolute measurements. We model its
first-order scaling as
\begin{equation}
    C_{\mathrm{mat}} \propto
    N_{\mathrm{served}} H
    \sum_g n_g c_g,
    \label{eq:feature_materialization}
\end{equation}
where $N_{\mathrm{served}}$ is the number of served ranking requests, $H$ is
the materialized history length, $n_g$ is the number of features in group $g$
per event, and $c_g$ is its empirically calibrated per-value materialization
compute on the common axis. These coefficients may differ across serving stacks.

For the common compute axis, we normalize full heterogeneous-event
materialization to one and empirically calibrate each schema. Semantic IDs
with CU embeddings fall in the low-single-digit
percentage range of this reference compute, while item features fall in the
high-single-digit range because they contain more values and a larger
floating-point payload. AMBER replaces both with a fixed-size cached token.
These ranges affect only horizontal compute coordinates; all NE values are
measured directly.

\section{Baseline Encoder Architectures}
\label{appendix:baselines}

\begin{figure}[htbp]
\centering
\begin{minipage}[b]{0.56\linewidth}
    \centering
    \includegraphics[width=\linewidth]{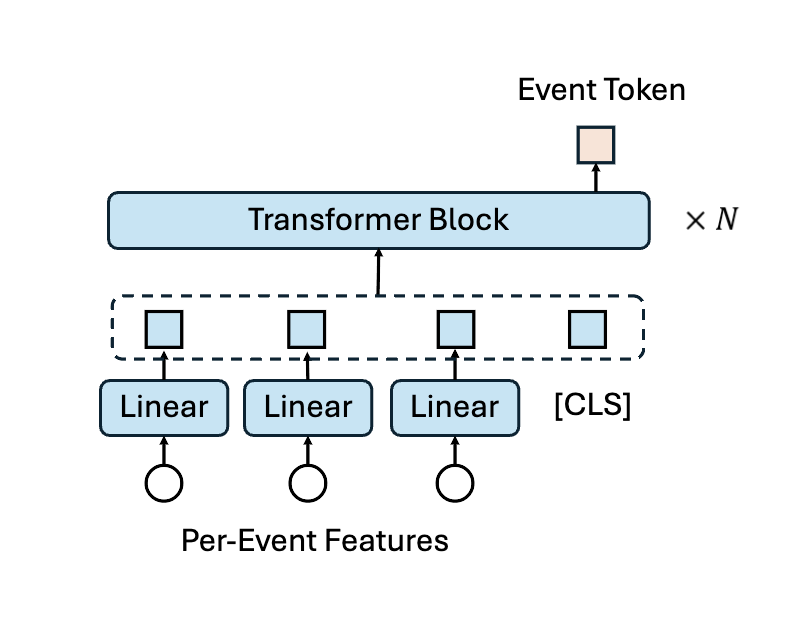}
    \centerline{(a) Event Tokenizer}
\end{minipage}
\hfill
\begin{minipage}[b]{0.40\linewidth}
    \centering
    \includegraphics[width=\linewidth]{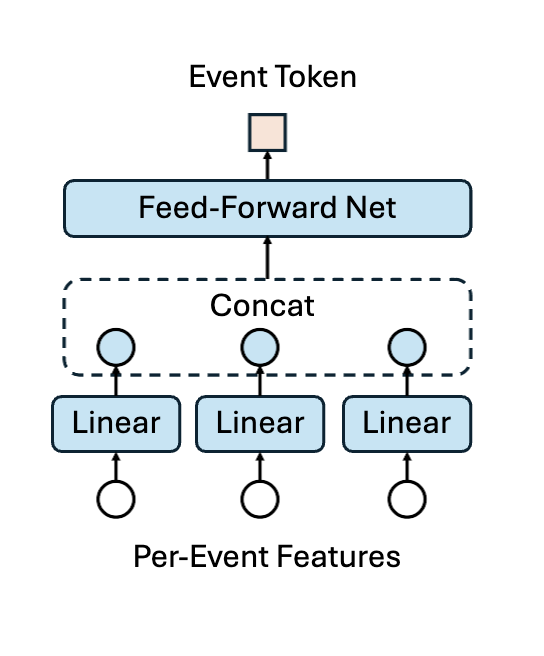}
    \centerline{(b) Concat+FFN}
\end{minipage}
\Description{Side-by-side comparison of the Event Tokenizer and Concat plus FFN architectures. The Event Tokenizer applies bidirectional Transformer blocks over projected feature tokens and learnable CLS tokens, whereas Concat plus FFN concatenates projected features and applies only a feed-forward network.}
\caption{Event encoder architectures. (a)~The Event Tokenizer processes projected per-event features and learnable \texttt{[CLS]} tokens with $N$ bidirectional Transformer layers, then maps the \texttt{[CLS]} output to an Event Token. (b)~Concat+FFN directly concatenates projected features and applies a feed-forward network without self-attention.}
\label{fig:encoder_architectures}
\end{figure}

Figure~\ref{fig:encoder_architectures} compares the Event Tokenizer with the Concat+FFN baseline used in the encoder architecture study (\S\ref{subsec:training_design}). Unlike the Event Tokenizer, Concat+FFN omits self-attention and limits inter-feature interactions to the feed-forward network.

\begin{table}[h]
\centering
\caption{Encoder architecture comparison. NE $\Delta$ is relative to the default four-layer bidirectional Transformer; positive values indicate degradation.}
\label{tab:encoder}
\begin{tabular}{lc}
\toprule
\textbf{Encoder} & \textbf{NE $\Delta\,\downarrow$} \\
\midrule
Bi-Transformer ($\times 4$) & --- \\
Bi-Transformer ($\times 2$) & $+0.09\%$ \\
Bi-Transformer ($\times 1$) & $+0.19\%$ \\
\midrule
DHEN ($\times 12$)~\citep{dhen} & $0.00\%$ \\
PMA ($\times 4$)~\citep{set_transformer} & $+0.20\%$ \\
Concat+FFN & $+0.22\%$ \\
\bottomrule
\end{tabular}
\end{table}

Table~\ref{tab:encoder} reports the comparison. DHEN ($\times 12$) matches the four-layer bidirectional Transformer at comparable FLOPs, while PMA and Concat+FFN trail by $0.20\%$ and $0.22\%$ NE, respectively. We select the Transformer because it is more flexible and achieves slightly higher practical efficiency through FlashAttention. Shrinking it to $\times 2$ and $\times 1$ increases NE by $0.09\%$ and $0.19\%$, respectively.

\section{HSTU-Style Input Baseline}
\label{appendix:hstulike}

AMBER's Event Tokenizer is agnostic to the downstream model: it can feed an HSTU-style sequential backbone as readily as the User LLM, and prior work finds these backbones perform comparably at scale~\citep{argus}. The informative comparison with HSTU is therefore over the \emph{input schema}, not the architecture. This ablation holds AMBER's User LLM backbone fixed and changes only the input schema to match HSTU's: each event retains only item-related categorical features, and user-side features are supplied once at the front of the sequence rather than encoded per event. To avoid handicapping it, we are generous with these user-side inputs, including all user float, embedding, and ID-list features (the reference implementation places only a few user categorical features at the front). This schema results in $1.0\%$ higher NE relative to AMBER ($+0.6\%$ vs.\ Incumbent (fair)).

\section{Pointwise Baseline}
\label{appendix:pointwise}

This baseline isolates the benefit of maintaining high snapshot resolution throughout history from the training-efficiency benefit of autoregressive modeling. Comparing AMBER directly with a conventional pointwise architecture would confound these effects because autoregressive training processes the full sequence in one pass, whereas pointwise training repeatedly encodes history for each impression. We therefore reproduce the pointwise input pattern within the same User LLM backbone and retain single-pass autoregressive training.

Each event is represented by three tokens: a high-resolution query token $A$ containing the full event features, an item-only history token $B$ containing the complete item-feature set, and an outcome token $C$. Although $B$ is lower-resolution than the full event, it is a strong upper bound whose feature count would still be computationally costly to materialize throughout the online history. The attention mask enforces the pointwise information flow. For event $i$, $A_i$ attends only to prior $B$ and $C$ tokens, so its prediction uses a high-resolution current query over an item-only history. Neither $B$ nor $C$ can attend to any $A$ token. This restriction prevents high-resolution information from entering $B$ or $C$ and propagating indirectly to future predictions; the persistent history therefore remains strictly item-only.

This construction keeps the backbone, optimization objective, compute budget, and autoregressive training efficiency matched to AMBER, while changing only whether high-resolution event information is retained in history. Pointwise improves over Incumbent (fair) by $0.1\%$ NE, isolating the benefit of autoregressive training, but remains $0.3\%$ behind AMBER. The remaining difference therefore supports the benefit of preserving high snapshot resolution throughout history rather than only at the current query.

\section{Robustness to Feature Materialization Latency}
\label{appendix:delay}

In realistic serving settings, a delay of 1--5 minutes between an event's occurrence and its feature materialization is typical. We experimented with simulating this delay during training by randomly masking recent events within a time window. Contrary to expectation, delay simulation degraded NE by $0.2\%$ even when evaluation accounted for the same offset, suggesting that the model is inherently robust to feature materialization latency and that the artificial masking removes useful training signal.

\section{Representation Drift Visualization}
\label{appendix:tsne}

\begin{figure}[htbp]
\centering
\begin{minipage}[t]{0.48\linewidth}
    \centering
    \includegraphics[width=\linewidth]{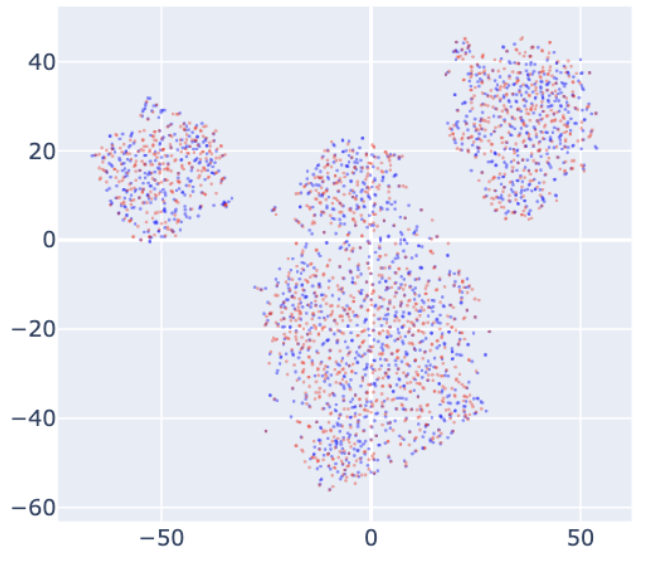}
    \centerline{(a) 1-day gap}
\end{minipage}
\hfill
\begin{minipage}[t]{0.48\linewidth}
    \centering
    \includegraphics[width=\linewidth]{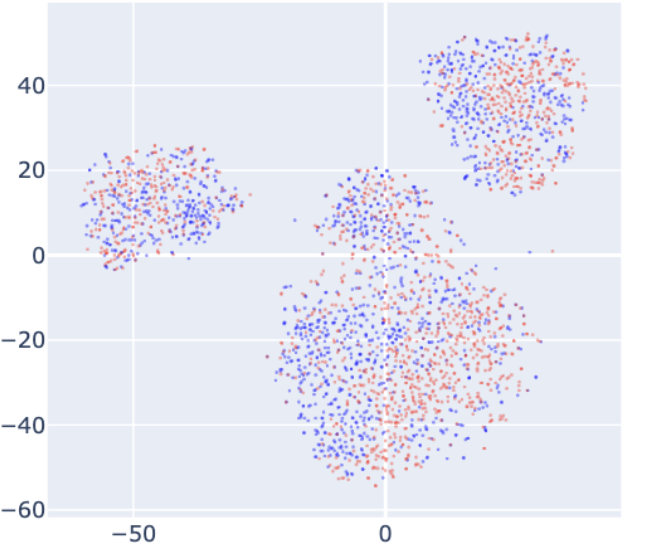}
    \centerline{(b) 20-day gap}
\end{minipage}
\Description{Two side-by-side t-SNE scatter plots compare Event Tokens generated for the same events by two encoder checkpoints. Points are colored by checkpoint. In the 1-day comparison, red and blue points overlap throughout several compact clusters. In the 20-day comparison, the overall cluster structure remains similar, but red and blue points separate within the clusters, indicating greater representation drift over the longer checkpoint interval.}
\caption{t-SNE visualization of Event Tokens from two encoder checkpoints (red vs.\ blue) encoding the same events. (a)~With a 1-day gap, the two distributions are well-mixed. (b)~With a 20-day gap, color separation emerges within clusters, indicating accumulated drift.}
\label{fig:tsne}
\end{figure}

\FloatBarrier
\section{Recurrent Training Stability}
\label{appendix:stability}

\begin{figure}[htbp]
\centering
\includegraphics[width=\linewidth]{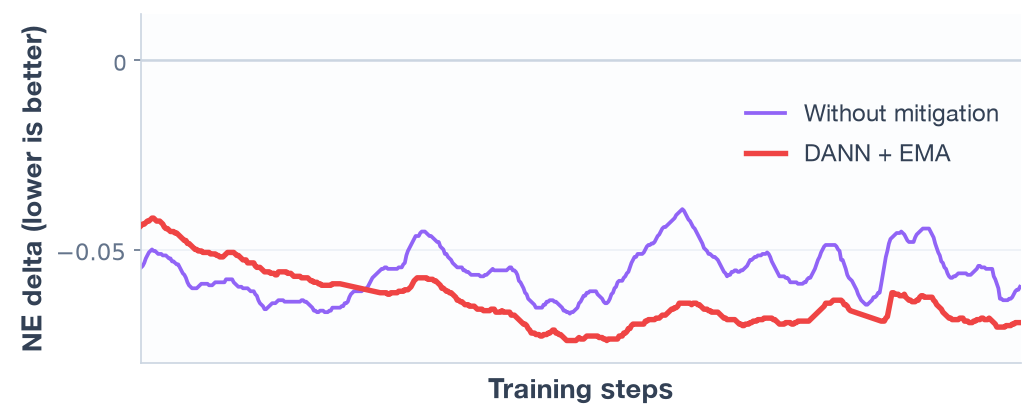}
\Description{NE trajectory comparison showing smoother training with DANN plus EMA versus without drift mitigation.}
\caption{NE trajectory during recurrent training. Without drift mitigation (purple), NE shows higher variance across training steps. With DANN + EMA (red), the smoother trajectory indicates more stable downstream predictions. Both runs use identical hyperparameters.}
\label{fig:ne_stability}
\end{figure}

\FloatBarrier
\section{Pre-trained Initialization Dynamics}
\label{appendix:pretrain}

\begin{figure}[H]
\centering
\includegraphics[width=\linewidth]{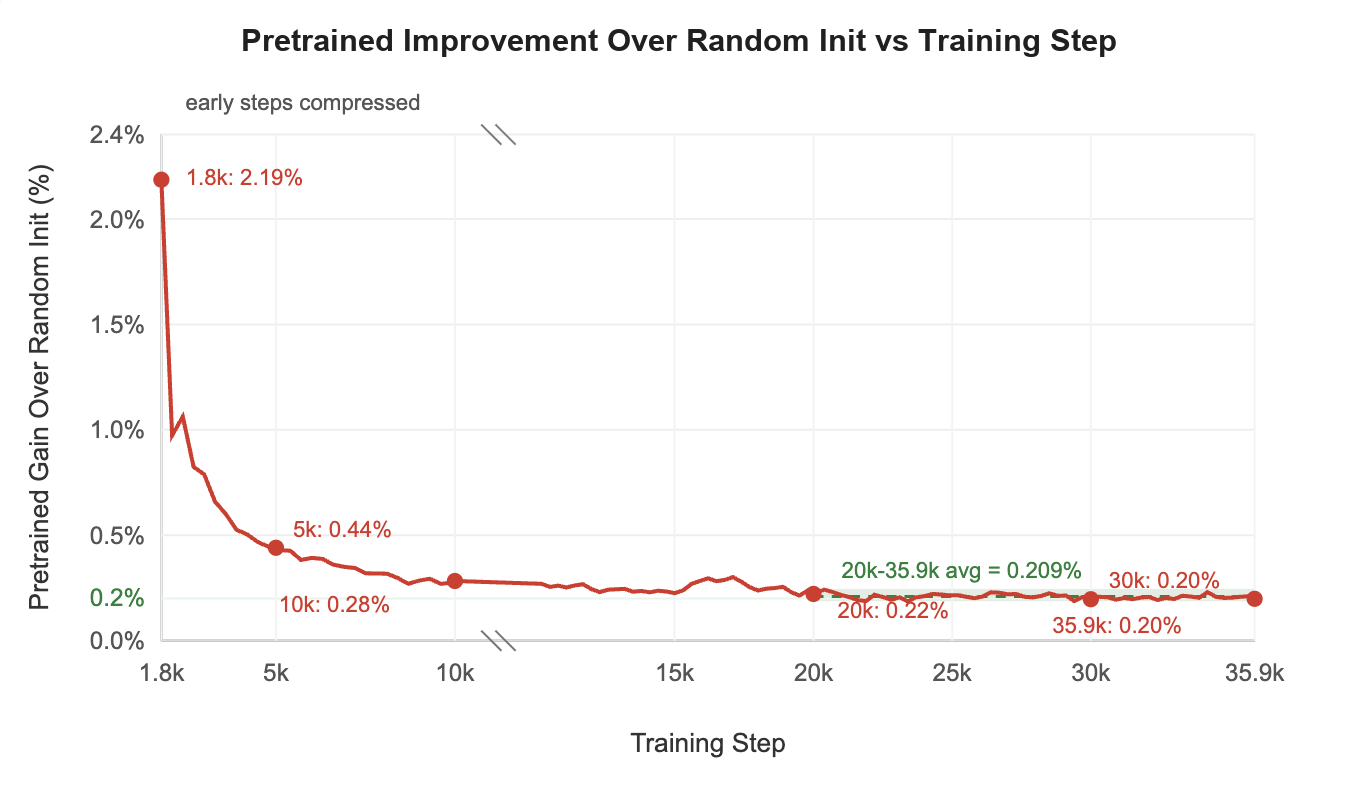}
\Description{Line plot of the relative NE advantage of pretrained over random LLM initialization versus training steps. The curve drops sharply from approximately 2.2 percent near the start of training, falls below 0.5 percent by roughly 10K steps, and then levels off near a persistent 0.2 percent advantage after 20K steps. A horizontal dashed line marks zero improvement.}
\caption{NE advantage of pre-trained LLM initialization over random initialization as a function of training steps. The advantage decays from ${\sim}2.2\%$ early in training to a stable ${\sim}0.2\%$ after 20K steps.}
\label{fig:pretrain}
\end{figure}

\FloatBarrier

\end{document}